\documentclass[12pt]{iopart}

\usepackage{iopams}  
\usepackage{color}
\usepackage{graphicx,subfigure}

\begin{document}

\title[Intra- and intercycle interference of angle-resolved electron emission...]
{Intra- and intercycle interference of angle-resolved electron emission in laser assisted XUV atomic ionization}

\author{A A Gramajo$^1$, R Della Picca$^1$, S D L\'opez$^2$ and D G Arb\'o$^2$}

\address{$^1$CONICET and Centro At\'omico Bariloche (CNEA), 8400 Bariloche, Argentina}
\address{$^2$ Institute for Astronomy and Space Physics IAFE
(CONICET-UBA), CC 67, Suc. 28, C1428ZAA, Buenos
Aires, Argentina}
\ead{gramajo.anaalicia@gmail.com}
\ead{diego@iafe.uba.ar}
\vspace{10pt}
\begin{indented}
\item[]\today
\end{indented}

\begin{abstract}
A theoretical study of ionization of the hydrogen atom due to an XUV pulse
in the presence of an IR laser is presented. 
Well-established theories are usually used to describe the %problem of 
laser assisted photoelectron effect. However, the well-known soft-photon
approximation firstly posed by Maquet \etal in Journal of Modern
Optics \textbf{54} 1847 (2007) and Kazansky's theory in Phys. Rev. A \textbf{%
82}, 033420 (2010) completely fails to predict the electron emission
prependicularly to the polarization direction. Making use of a %simple 
semiclassical model, 
we study the angle-resolved energy distribution of photoelectrons
for the case that both fields are linearly polarized in the
same direction. We
thoroughly analize and characterize two different emission regions in the
angle-energy domain: (i) the parallel-like region with contribution of two
classical trajectories per optical cycle and (ii) the perpendicular-like
region with contribution of four classical trajectories per optical cycle. We show that our semiclassical
model is able to asses the interference patterns of the angle-resolved
photoelectron spectrum in the two different mentioned regions. Electron
trajectories stemming from different optical laser cycles give rise to
angle-independent \textit{intercycle} interference known as
sidebands. These sidebands are modulated by an angle-dependent
coarse-grained structure coming from the \textit{intracycle} interference of
the electron trajectories born during the same optical cycle. We show
the accuracy of our semiclassical model as a function of the time delay
between the IR and the XUV pulses and also as a function of the laser
intensity by comparing
the semiclassical predictions of the angle-resolved photoelectron spectrum
with the continuum-distorted wave strong field approximation and the \textit{%
ab initio} solution of the time dependent Schr\"{o}dinger equation
\end{abstract}

%\noindent{\it Keywords}: PHOTOIONIZATION, LAPE, INTRACYCLE, INTERCYCLE, ANGLE RESOLVED 

\submitto{\JPB}
\maketitle
%\twocolumn

%----------------------------------------------
\section{Introduction}
%----------------------------------------------

Most of the experiments on laser assisted photoelectric effect (LAPE)
combined a fundamental quasimonochromatic laser (IR) with its high-order
harmonic product acting both on rare-gas atoms (XUV and soft-X-ray
radiations) \cite{Veniard95,Schins96}. Lately, new sources produced from
X-ray free-electron laser (XFEL) in the strong field regime were used to
achieve multi-photon spectroscopy involving synchronized IR and XUV pulses 
\cite{Meyer08,Meyer10a,Radcliffe12,Mazza14,Hayden16,Duesterer16}. The
photoelectron (PE)  spectra from rare gas atoms have been extensively studied in
the simultaneous presence of the two pulses --the XUV and IR laser-- with a
time-controlled delay working as a pump-probe experiment \cite%
{Drescher05,Goulielmakis04,Johnsson05}. 
Whereas first experiments measured
the angle-integrated photoelectron emission, only recently, simultaneous energy- and angle-resolved PE spectra 
%angle-resolved photoelectron spectra 
have been gauged with high degree of resolution 
\cite%{Meyer10a,
{Duesterer16,Guyetand05,Guyetand08,Bordas04,%Meyer10b,
Meyer12,Duesterer13,Picard14}.
The determination
of angle-resolved photoelectron spectra requires state of the art %sophisticated 
techniques employing several electron time-of-flight (TOF) analyzers mounted at different angles \cite{Duesterer16,Meyer12},  cold target recoil ion momentum spectroscopy (COLTRIMS) \cite{Picard14} or velocity map imaging (VMI) techniques \cite{Guyetand08,Bordas04,Duesterer13}.
Depending on the XUV pulse duration ($\tau_X$), two well-known regimes --sideband and
streaking-- has been distinguished \cite%
{Duesterer13,Itatani02,Fruehling09,Wickenhauser06,Nagele11}. In the former,
where the XUV pulse is longer than the laser period ($T_L$), the photoelectron
energy spectrum shows a main line associated with the absorption of one XUV
photon accompanied by sideband lines associated with additional exchange of
laser photons \cite{Schins96,OKeeffe04,Glover96,Aseyev03,Guyetand05}. In the
latter, as the XUV pulse is much shorter than the % inverse of the 
laser wavelength, %frequency, 
the electron behaves like a classical particle getting linear
momentum from the IR laser field at the instant of ionization \cite%
{Drescher05,Wickenhauser06,Fruehling09,Swoboda09,Nagele11}. The analysis of
the resulting two-color photoelectron spectra can provide information about
the high-frequency pulse duration, laser intensity, and the time delay
between the two pulses. Moreover, the duration of atomic transitions, like
the Auger decay, has been measured with unprecedented levels of
accuracy in the attosecond realm \cite{Schins94,Kazansky10a,Meyer12}.

Precise calculations of the response of a rare gas atom are based on quantum
mechanical concepts, i.e., by solving \textit{ab initio} the time dependent
Schr\"{o}dinger equation (TDSE) for the atomic system in the presence of the
two pulses within the dipole approximation. The numerical resolution of the
TDSE for a multi-electron system relies on the single-active electron
approximation with model potentials that reproduce the bound state spectrum
of the atom with satisfactory accuracy \cite{Nandor99,Muller99}. Simplified
theories are also very useful at the time of understanding the physical
processes involved in LAPE. Most of the approximated models of LAPE
processes are based on the strong field approximation (SFA) \cite%
{Kazansky10a,Kazansky10b, Bivona10}. For example, the broadly used
soft-photon approximation (SPA) \cite{Maquet07,Jimenez13} provides a useful
description of some general features in experiments \cite{Meyer06, Meyer08,
Meyer10a,OKeeffe04,Duesterer16,Hayden16}, however, it completely fails to
reproduce the measured electron yield from $s$-bound states with high
emission angles, predicting no contribution in the direction perpendicular
to the polarization axis in LAPE \cite{Maquet07,Jimenez13,Taieb08,Kroll73},
contrarily to TDSE calculations. Besides, the analytic angle-resolved PE
spectra derived by Kazansky \etal \cite{Kazansky10a,Kazansky10b}
and Bivona \etal \cite{Bivona10} are based on simplifications of
the temporal integration within the SFA. Following in Bivona \etal
footsteps \cite{Bivona10}, in previous works we have presented a
semiclassical approach that describes the XUV+IR multiphoton ionization with
emission parallel and perpendicular to the polarization direction of both
fields \cite{Gramajo16,Gramajo17}. Within a one-dimensional semiclassical
model (SCM), the PE spectrum was interpreted as the coherent superposition
of electron trajectories emitted during the action of the XUV pulse, giving
rise to intra- and intercycle interference patterns \cite{Arbo10a,Arbo10b,ARBO2012}.
As far as we know, LAPE ionization has not been studied successfully in
detail for arbitrary emission directions. The poor agreement between
theoretical and experimental PE angular distributions for the two-color
above threshold ionization leads to the necessity of a more comprehensive
theoretical description \cite{Haber09,Haber10}.

In this paper we extend the one-dimensional semiclassical approximation (for
parallel \cite{Gramajo16} and perpendicular emission \cite{Gramajo17})
towards the analysis of the angle-resolved laser assisted photoemission
spectra of hydrogen atoms by an XUV pulse in the intermediate case between
the sideband and streaking regimes, i.e., $\tau _{X}\gtrsim T_{L}$. We
characterize different regions of the energy-angle plane with different
number of contributing electron trajectories coherently superimposed to form
the interference pattern. Our SCM leads to a simple analytical expression of
the doubly differential energy-angle distribution similar to the equation of
the diffraction grating in the time domain giving rise to intercycle
interferences (sidebands) modulated by the intracycle pattern (also known as
the gross structure \cite{Kazansky10a}). We show that our SCM reproduces the
sidebands very accurately (compared to SFA and TDSE computations) for all
emission angles, even for directions close to perpendicular emission, where
Kazansky's theory \cite{Kazansky10a,Kazansky10b} and the SPA \cite{Maquet07,Jimenez13} break down. Besides, we show that
%for high intensity lasers 
the SCM also predicts the downshift of the energy
of the continuum states %of the atom 
by the ponderomotive energy $U_{p}$ due
to the average wiggling of the electron driven by the laser field.
%that it comes relevant for high intensity lasers.

The paper is organized as follows. In Sec. \ref{theory} we describe the SCM
for the case of laser assisted XUV ionization emphasizing the
characterization of the electron trajectory contributions in the different
regions of the energy-angle domain. 
Details of the theoretical calculation are developed in the Appendix.
In Sec. \ref{results}, we present the
results and discuss over their comparison with the corresponding calculations within
the \textit{ab initio}  TDSE and the SFA. Concluding remarks are presented in
Sec. \ref{conc}. Atomic units are used throughout the paper, except when
otherwise stated.

%----------------------------------------------
\section{\label{theory}Theory of the semiclassical model}
%----------------------------------------------

We consider the ionization of an atomic system by the interaction with an extreme ultra violet (XUV) finite laser pulse assisted by an infra red (IR) laser, both linearly polarized in the same direction $\mathbf{e}_{z}$.
In the single-active-electron (SAE) approximation the time-dependent Schr\"{o}dinger equation (TDSE) reads
\begin{eqnarray}
\rmi\frac{\rmd}{\rmd t}\left\vert \psi (t)\right\rangle =
\Big[ H_0  + H_{int}(t)  \Big]
\left\vert
\psi (t)\right\rangle , 
\label{TDSE}
\end{eqnarray}%
where $H_0=\mathbf{p}^2/2+V(r)$ is the time-independent atomic Hamiltonian, whose first term
corresponds to the electron kinetic energy, and its second term to the electron-core Coulomb interaction.
The second term in the right-hand side of equation (\ref{TDSE}), i.e,  
$H_{int}=\mathbf{r}\cdot \mathbf{F}_{X}(t) + \mathbf{r}\cdot \mathbf{F}_{L}(t)$,
stands for the interaction of the atom with both time-dependent XUV [$\mathbf{F}_{X}(t)$]
and IR [$\mathbf{F}_{L}(t)$] electric fields in the length gauge.

As a consequence of the interaction, the bound electron in the initial atomic state 
$|\phi _{i}\rangle $ is emitted with momentum $\mathbf{k}$ and energy $%
E=k^{2}/2 $ into the final unperturbed state $|\phi _{f}\rangle $. The
photoelectron momentum distributions can be calculated as 
\begin{equation}
\frac{\rmd P}{\rmd\mathbf{k}}\mathbf{=}\left\vert T_{if}\right\vert ^{2},  \label{P}
\end{equation}%
where $T_{if}$ is the T-matrix element corresponding to the transition $\phi
_{i}\rightarrow \phi _{f}$. 

Within the time-dependent distorted wave theory, the transition amplitude in the prior form and length gauge is expressed as \cite{DellaPicca13,Macri03}
\begin{equation}
T_{if}= -\rmi\int_{-\infty}^{+\infty}\rmd t \,\langle\chi_{f}^{-}(\mathbf{r},t)|H_{int}(\mathbf{r},t)|\phi_{i}(\mathbf{r},t)\rangle, 
\label{Tif}
\end{equation}
where $\phi_{i}(\mathbf{r},t)=\varphi_{i}(\mathbf{r})e^{\rmi I_{p} t}$ is the initial atomic state,
$I_{p}$ the ionization potential, and $\chi_{f}^{-}(\mathbf{r},t)$ is the distorted final state.  The SFA neglects the Coulomb core-electron interaction in the final channel, therefore we use the well known Volkov wavefunction \cite{Volkov} to represent the free electron in the electromagnetic field. The Volkov wavefunction reads
\begin{equation}
\chi_{f}^{V}(\mathbf{r},t)
=(2\pi)^{-3/2}\, \exp{\left[\rmi\big(\mathbf{k}+\mathbf{A}(t)\big)\cdot\mathbf{r}+
\frac{\rmi}{2}\int_{t}^{\infty}\rmd t'\big(\mathbf{k}+\mathbf{A}(t')\big)^2\right]},
\label{Volkov}
\end{equation}
and the vector potential due to the total external field is defined as 
$\mathbf{A}(t) = -\int_0^t \rmd t'[\mathbf{F}_{X}(t')+\mathbf{F}_{L}(t')]$. 
In sec. 3, we will denote the SFA as the numerical integration of the transition matrix in equation (\ref{Tif}) by including the Volkov phase [equation (\ref{Volkov})] in the final channel.

With the appropriate choice of the IR and XUV laser parameters considered, we can assume
that the energy domain of the LAPE processes is well separated from the 
IR ionization one. 
In other words, the
contribution of IR ionization is negligible in the energy domain where the absorption of one XUV photon takes place. 
Furthermore, considering the rotating wave approximation we can consider (neglect) the absorption (emission) of an XUV photon and, thus, the expression of the linearly polarized XUV pulse is reduced to 
$\mathbf{F}_{X}(t)\sim F_{X0}(t)\exp{(-\rmi \omega _{X}t)}\mathbf{e}_{z}/2$, where $\omega_{X}$ is the XUV field frecuency. Finally, equation (\ref{Tif}) can be written as:
\begin{equation}
T_{if}=-\frac{\rmi}{2}\int_{-\infty }^{+\infty }\rmd t\ d_z%^{\ast }%
\left( \mathbf{k}+\mathbf{A}(t)\right) F_{X0}(t)\, \exp {\left[\rmi S(t)\right]} ,
\label{Tifg}
\end{equation}%
where the dipole element 
$\mathbf{d}(\mathbf{v})$ is given by 
\begin{equation}
\mathbf{d}%^{\ast }
(\mathbf{v})=\frac{1}{(2\pi )^{3/2}}\int \rmd\mathbf{r}\exp {[-}\rmi \mathbf{v%
}\cdot \mathbf{r}{]\ }\mathbf{r}\ \varphi _{i}(\mathbf{r}),
\label{dipole}
\end{equation}
and the generalized action is
\begin{equation}
S(t)=-\int_{t}^{\infty }\rmd t^{\prime }\left[ \frac{\left( \mathbf{k}+\mathbf{A}%
(t^{\prime })\right) ^{2}}{2}+I_{p}-\omega _{X}\right].  \label{action}
\end{equation}

As $\omega_X$ is much higher than the frequency of the IR pulse $\omega_{L}$  (and for XUV pulses weaker than the IR laser), we
can consider the vector potential as due to the laser field only, 
neglecting its XUV contribution \cite{DellaPicca13,Nagele11,Gramajo16,Gramajo17}.
Hence  the total vector potential can be written as $\mathbf{A}(t)\simeq \mathbf{A}_{L}(t) = A_{L0} \sin (\omega_L t) \mathbf{e}_{z}$
 since  during the temporal lapse when the XUV pulse is acting the IR electric field is modeled as a cosine-like wave. Here  $A_{L0} = F_{L0} /\omega_L $ and $F_{L0}$ is the amplitude of the IR laser field.

For simplicity, we consider a hydrogen atom initially in the ground state and we restrict our analysis to the case where the XUV pulse duration is a multiple of the laser optical cycle, \textit{i.e.} $\tau_X = N T_L$ with $N=1,2,...$ and $T_L=2\pi/\omega_L$. Since both fields are linearly polarized in $\mathbf{e}_{z}$, we describe the  photoelectron momentum in cylindrical coordinates as: $\mathbf{k}=k_{z}\mathbf{e}_{z}+ k_{\perp}\mathbf{e}_{\perp}$.

\begin{figure*}[t!]
\subfigure[]{\includegraphics[width=0.5\textwidth]{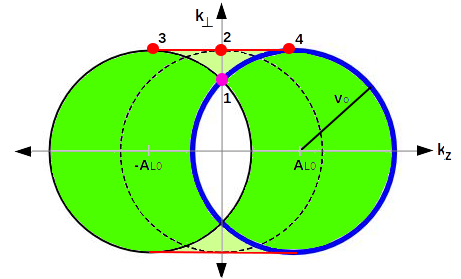}}
\subfigure[]{\includegraphics[width=0.5\textwidth]{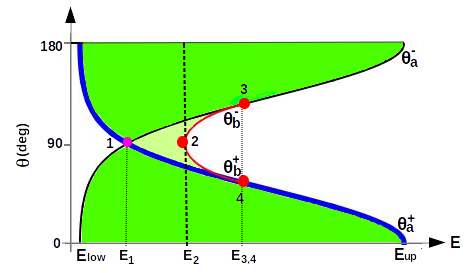}}
\caption{Color online. (a) Schematic picture for the classically allowed region in
the momentum space $k_{z}\mathbf{e}_{z} +k_{\perp}\mathbf{e}_{\perp}$. The dash-dot circle of radius $v_0$
centered at origin represents the main emission line (due to XUV ionization
without laser field). As the laser vector potential oscillates, the circle
shifts horizontally by $-A_{L}(t)$ with amplitude $A_{L0}$. In the
parallel-like region (green) there are two classical electron trajectories
contributing to the probability distribution. In the perpendicular-like
region (light green) the number of contributing trajectories is four. The
white area represents the classically forbidden region. (b) The same as in
(a) but in the energy-angle domain.}
\label{figure1}
\end{figure*}

The SCM approach consists of solving the time integral of  equation \eref{Tifg} by means of the saddle-point approximation \cite{Chirila05,Corkum94,Ivanov95,Lewenstein95}. 
In this sense, the transition
probability can be written as a coherent superposition
of the amplitudes of all electron classical trajectories with final momentum $\mathbf{k}$ 
over the stationary points $t_{s}$ of the generalized action $S(t)$:
\begin{equation}
T_{if}=\sum_{t_{s}}\frac{\sqrt{2\pi }F_{X0}(t_{s})d_{z}
(\mathbf{k}+\mathbf{
A}_{L}{(t}_{s}{)})}{\left\vert \ddot{S}(t_{s})\right\vert ^{1/2}}\exp \left[\rmi S(t_{s})%
+ \rmi \frac{\pi}{4} \mathrm{sgn} [ \ddot{S}(t_{s})] \right],
\label{Tifsaddle}
\end{equation}%
where 
 $\ddot{S}(t_{s}) = -\left[ \mathbf{k}+\mathbf{A}_{L}(t_s) \right]%
\cdot \mathbf{F}_{L}(t_s)$, $\mathrm{sgn}$ denotes the sign function,
 and the dipole element from the $1s$ reads
\begin{equation}
d_{z}(\mathbf{v})=-\frac{\rmi 2^{7/2}}{\pi}(2I_{p})^{5/4}\frac{\mathbf{e}_{z}\cdot\mathbf{v}}{[\mathbf{v}^2 + 2I_{p}]^3} \,\, .
\label{dip}
\end{equation}
The ionization times $t_{s}$ fulfill the equation $\rmd S(t)/\rmd t|_{t=t_s}=0$, \textit{i.e.},
\begin{equation} 
(k_{z}+A_{L}(t))^2+k_{\perp}^2 =  v_{0}^{2}, \label{circle}
\end{equation}
where $v_{0}=\sqrt{2(\omega_{X}-I_{p})}$. 
In the momentum space the equation (\ref{circle}) is the circumference with center at $-A_{L}(t)\mathbf{e}_{z}$ and radius $v_{0}$. The center position of the circumference oscillates with the time-dependent vector potential.
In figure \ref{figure1}(a) we show the representation of equation (\ref{circle}). At time zero the circumference  plotted with dashed line is centered at the origin and starts to move to left as the potential vector increases. When the potential vector reaches the maximal amplitude $A_{L0}$ the circumference is situated at the left with center at $-A_{L0}\mathbf{e}_{z}$, then it moves to the right. At the end of the IR cycle it returns to the origin. 
The shaded area indicates the $\mathbf{k}$ values that were reached  by the circumference described by equation (\ref{circle}) at some time during one IR cycle. 
In other words, the classically allowed momenta are all points of the shaded area  for which there exists a time $t_s$ that verifies equation (\ref{circle}).
Outside this domain, ionization times are complex 
giving rise to non-classical trajectories with  
exponentially decaying factors and thus  minor relevance compared to real ones. 
Hereinafter, we restrict our SCM to classical allowed momenta. 

In view of the following analysis, we can distinguish two regions in figure \ref{figure1}: the parallel-like region (in green) and  the perpendicular-like one (light green) that is delimited by the points 1 to 4. 
In each region, the SCM amplitude is derived analogously 
to previously studied parallel and perpendicular cases \cite{Gramajo16,Gramajo17}.

%-------FIGURE 1----------------------------------------------------------
%-------FIGURE 1----------------------------------------------------------

Alternatively, these regions can be also identified in the energy-angle plane \textit{via} the transformation $(k_{z}, k_{\perp}) \rightarrow (E, \theta)$, with  
$ E= (k_{z}^2+k_{\perp}^2)/2$ and 
$\tan{\theta}= k_{\perp}/k_{z}$ [see figure \ref{figure1}(b)]. 
In this plane, the curves delimiting the allowed regions are defined by 
\begin{eqnarray}
\theta^{\pm}_{a}(E)&=&\arccos{\left( \pm\ \frac{ E+ A_{L0}^{2}/2- v_{0}^{2}/2}{ \sqrt{2E}A_{L0}} \right)}, \label{theta1}\\
\theta^{\pm}_{b}(E)&=&\arcsin{ \left(\pm\frac{v_{0}}{\sqrt{2E}}\right)} + \frac{(1\mp 1)\pi}{2}, \label{theta2}
\end{eqnarray}
which, in momentum space, correspond to the circles in thick blue, thin black lines and the connecting points 3-4 in thin red (grey) line, respectively. 
The lower and upper classical values for the electron energy are $\theta$ depending. 
For example, in the forward and perpendicular emission cases ($\theta=0^o$ and $90^o$ respectively), these values are 
$E_\textrm{\tiny low,up}= (v_{0}\mp A_{L0})^2/2$ 
and
%whereas in the perpendicular emission case %($\theta=90^o$) 
%the classical boundaries are situated at 
$E_{1}=(v_{0}^2-A_{L0}^2)/2$, $E_{2}=v_{0}^2/2$ for the perpendicular emission case, in agreement with previous works \cite{Gramajo16,Gramajo17}.

The deduction of the analytical expressions for the ionization times $t_{s}$ that fulfill equation \eref{circle} is detailed in the Appendix.
After a bit of algebra     
%Finally, from \Eref{Tifsaddle}, 
%and in same way  as in  \cite{Gramajo16,Gramajo17} 
%(see the Appendix for more details), 
it can be shown that the emission probability in equation (\ref{Tifsaddle}) is 
\begin{eqnarray}
|T_{if}|^{2}&=4\Gamma{(k_{\perp})}F(\mathbf{k})B(k), \label{Tif4}
\end{eqnarray}
%
%where $k=|\mathbf{k}|$, $
where
\begin{equation}
B(k)=\frac{\sin^{2}(N\tilde{S}/2)}{\sin^{2}(\tilde{S}/2)},
\end{equation}
\begin{equation}
\tilde{S} = (2\pi/\omega_{L} )(E + I_{p} + U_{p}-\omega_{X}),
\label{Smonio}
\end{equation}
 and the \textit{intracycle} factor is
\begin{eqnarray}
F(\mathbf{k})&=
\left| 
f_+(\mathbf{k})
\cos{  \left(\frac{\Delta S_+ }{2} +  \frac{\pi}{4} \mathrm{sgn}[\beta_+(\mathbf{k})]  \right) } 
 -
 \Theta{  \scriptstyle \left(1+\frac{\beta_-}{A_{L0}}\right)}
f_-(\mathbf{k})
\cos{\left(\frac{\Delta{S}_{-}}{2}
-  \frac{\pi}{4}\mathrm{sgn}[\beta_{-}(\mathbf{k})] \right) }
\right|^{2},  \nonumber \\
\label{F}
\end{eqnarray}
where $\beta_{\pm}(\mathbf{k}), f_\pm(\mathbf{k})$ and  $\Delta{S}_{\pm}$ are defined in equations \eref{betamm}, (\ref{ff}) and (\ref{DeltaSpara}) of the appendix, respectively, and $\Theta$ is the Heaviside function.
The ionization rate $\Gamma(k_{\perp})$  in equation (\ref{Tif4}) is identical for all subsequent ionization trajectories which depend on the perpendicular component of the final momenta $k_{\perp}$, i.e., 
\begin{eqnarray}
 \Gamma(k_{\perp})&=\frac{4F_{X0}^{2}}{\pi F_{L0}\omega_{X}^{6}}\sqrt{v_{0}^{2}-k_{\perp}^{2}}. \label{gama}
\end{eqnarray}

The equation (\ref{Tif4}) indicates that the PE spectrum can be factorized in two different contributions: (i) the \textit{intracycle interference} stemming from trajectories within the same cycle governed by the factor $F(\mathbf{k})$, and (ii) the \textit{intercycle interference} stemming
from trajectories released at different cycles, resulting in the well-known sidebands given by the factor $B(k)$. The latter factor is periodic in the final photoelectron energy with peaks at positions
\begin{equation}
 E_{n}=\omega_{X} + n\omega_{L} - I_{p}- U_{p},
\label{energy}
\end {equation}
where $n=0,\pm 1,\pm 2, ....$ is interpreted as the number of IR photons absorbed ($n>0$) or emitted ($n<0$), added to the absorption of one XUV photon and downshifted by the ponderomotive energy $U_{p}$. 
When the duration of the pulses extends infinitely we have $B(k) \rightarrow \sum_n \delta(E-E_n)$, which stands for the conservation of energy.

%-------FIGURE 3----------------------------------------------------------
\begin{figure}[h]
  \centering
    \includegraphics[width=0.7\textwidth]{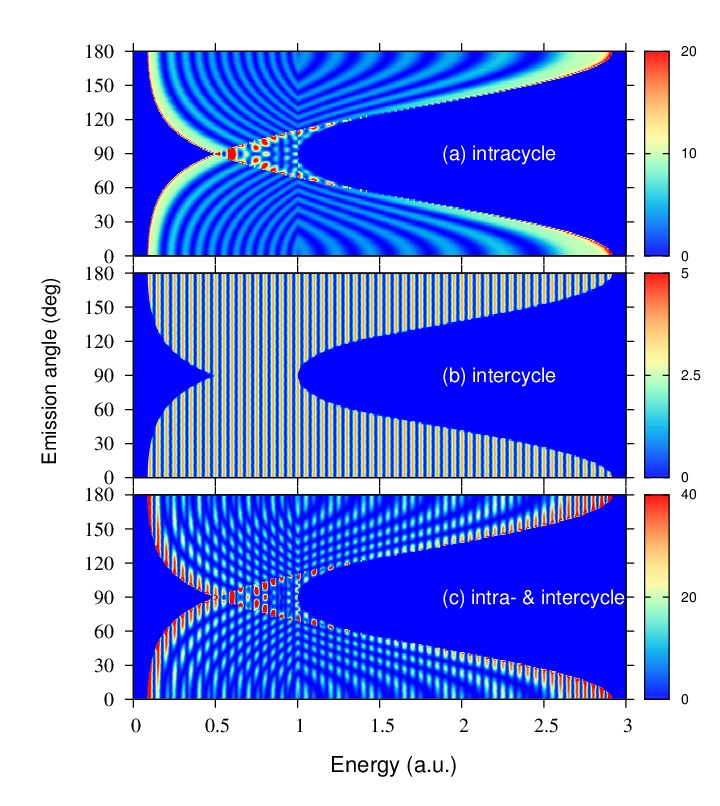}
  \caption{(a) SCM intracycle factor $F(\mathbf{k})$, (b) SCM intercycle
interference factor $B(k)$ considering $N=2$ optical cycles, and (c)
the product $F(\mathbf{k}) B(k)$ showing the interplay of inter- and intracycle
interferences. 
The IR and XUV laser parameters are $\omega_{L}=F_{L0}=F_{X0}=0.05$ a.u. and
$\omega_{X}=1.5$ a.u.}
  \label{figure3}
\end{figure}
%-------FIGURE 3----------------------------------------------------------

In figure \ref{figure3}, we show the respective contributions of intra- and intercycle factors, 
$F(\mathbf{k})$ and $B(k)$,
to the SCM emission probability in equation (\ref{Tif4}) for a  XUV pulse duration of $2T_{L}$.      
In \Fref{figure3}(a) we present the intracycle factor $F(\mathbf{k})$ that depends on both photoelectron energy and angle. We see that %the factor $F(\mathbf{k})$ [equation (\ref{F})] 
this factor has a richer structure in the perpendicular-like region (four contributing electron trajectories) than in the parallel-like region (two contributing trajectories).
Furthermore,
the equation \eref{F} predicts a jump due to the discontinuity of the function $\mathrm{sgn}$.
Since the sign of $\beta_-(\mathbf{k})$ is constant throughout the domain, we only expect
a discontinuity in the intracycle factor at $\beta_+=0$ (when its sign changes), that it can
be recognized in the figure at $E_\textrm{\scriptsize{disc}}=v_0^2/2 =1$ independently of the emission angle.
In the next section we analyze this discontinuity as a function of the beginning time of the XUV pulse.

In \Fref{figure3}(b) we plot the intercycle factor $B(k)$ in the classical domain, we observe the periodic stripes separated by $\omega_L$ at energies $E_{n}$ in according to equation (\ref{energy}). 
Finally, when both intra- and intercycle factors are multiplied, we obtain the spectra plotted in \Fref{figure3}(c). 
We observe that the intracycle interference
pattern works as a modulation  of the intercycle interference pattern. 
The agreement  between  present semiclassical  description and  ones obtained by 
SFA and TDSE  
is discussed in the next section.

%----------particularizacion a 0 y 90 grados

Finally, it is interesting to note that the present angle-dependent SCM formalism comprises the forward and perpendicular emission as particular cases, which were already analyzed in our previous works \cite{Gramajo16,Gramajo17}.
In fact,  when the electron emission is parallel to both laser fields ($k_\perp=0$), the second term inside the square modulus of $F(\mathbf{k})$ [equation (\ref{F})] is null and, thus,  equation (\ref{Tif4}) becomes equation (23) of \cite{Gramajo16}.
On the other hand, in the perpendicular electron emission case ($k_z=0$) we
have that $\beta_{-} = -\beta_{+}$ and then $\Delta{S}_{-}=\Delta{S}_{+}\equiv \Delta{S}$.
Thus the expression (\ref{F}) can be rewritten as [see also equation 18 of \cite{Gramajo17}]
 \begin{equation}
F(\mathbf{k}) = \frac{4}{ \scriptstyle \sqrt{1-\beta^2/A_{L0}^2}}
              \underbrace{  \cos^2 \left(\frac{\Delta{S}}{2}+\frac{\pi}{4}\right) }_\textrm{ \scriptsize  intra-half-cycle}
                \sin^2 \left(\frac{\tilde{S}}{4}\right),
\end{equation}
which can be understood as the contribution of an intra-half-cycle factor and $\sin^2(\tilde{S}/4)$ that 
interferes destructively for the absorption and/or emission of an even number of IR photons, 
which leads to the exchange of only an odd number of laser photons in the formation of the
sidebands. This fact can be observed in \Fref{figure3}(c) where even sideband peaks are canceled at $\theta=90^\circ$. 
%

%--------------------------------------------------------------
\section{Results and discussion}
%--------------------------------------------------------------
\label{results}

In the following, we analyze the angle resolved PE spectrum
\begin{equation}
\frac{\rmd^2 P}{\sin{\theta}\rmd{E}\rmd{\theta} }=2\pi k|T_{if}|^{2}
\label{dP}
\end{equation}
and 
compare the outcome of the SCM with quantum
calculations within the SFA \cite%
{Arbo08a,Kazansky10a,Kazansky10b,Bivona10,Arbo10a,Arbo10b} and by solving
the TDSE \cite{Tong97,Tong00,Tong05}.  
We model the XUV and IR laser pulses as 
\begin{equation}
\mathbf{F}_{i}(t)=F_{i0}(t-t_{ib})\ \cos \left[ \omega _{i}\left( t-\Delta _{i}-%
\frac{\tau _{L}}{2}\right) \right]\mathbf{e}_{z},  \label{i-field}
\end{equation}%
where $i=$ L and X denotes the IR laser and XUV pulses, respectively. The
envelopes of the electric fields in equation (\ref{i-field}) were chosen with a
trapezoidal shape comprising one-cycle ramp on and one-cycle ramp off, i.e.,%
\begin{equation*}
F_{i0}(t)=F_{i0}\left\{ 
\begin{array}{ccc}
t/T_{i} & \mathrm{if} & 0\leq t\leq T_{i} \\ 
1 & \mathrm{if} & T_{i}\leq t\leq \tau _{i}-T_{i} \\ 
(\tau _{i}-t)/T_{i} & \mathrm{if} & \tau _{i}-T_{i}\leq t\leq \tau _{i}%
\end{array}%
\right. 
\end{equation*}%
and zero otherwise, where $T_{i}=2\pi /\omega _{i}$ and $\tau _{i}$ are the $%
i-$field period and pulse duration, respectively. For the sake of
simplicity, we suppose that the duration of both laser fields comprises
integer number of cycles, i.e., $\tau _{i}=N_{i}T_{i}$ where $N_{i}$ is a
natural number. We choose the origin of the time scale as the beginning of
the IR laser pulse, i.e., $t_{Lb}=0$, with no displacement of it %the laser pulse 
$\Delta _{L}=0,$ so that the IR laser field is a cosine-like pulse
centered in the middle of the pulse. 
The beginning time of the XUV pulse $t_{Xb}=\Delta_{X}+\tau _{L}/2-\tau _{X}/2$ marks the starting time of the active window for LAPE, that corresponds to the temporal interval $[t_{Xb}, t_{Xb}+\tau_X]$ when both pulses are superimposed.

\begin{figure}[h]
  \centering
    \includegraphics[width=0.7\textwidth]{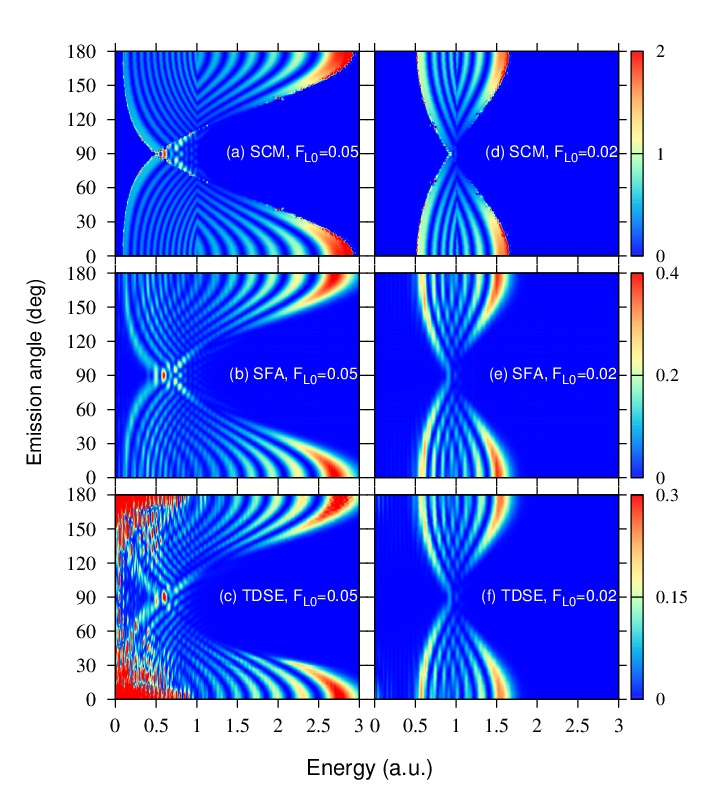}
  \caption{Angle-resolved photoelectron spectra in arbitrary units for an XUV pulse
duration of $\tau_{X}=T_{L}$ and time delay $\Delta_{X}=0$, calculated at different laser peak fields ($%
F_{L0}=0.05$ a.u. in a, b, and c, and $F_{L0}=0.02$ a.u. in d, e, and f)
within the SCM (a and d), the SFA (b and e) and the TDSE (c and f). The IR
laser frequency is $\omega_{L}=0.05$ a.u. and the XUV pulse parameters are $%
\omega_{X}=1.5$ a.u. and $F_{X0}=0.05$ a.u..}
  \label{figure4}
\end{figure}

Hereinafter, in our calculations we use the IR and XUV pulses with
frequencies as $\omega _{L}=0.05$ and $\omega _{X}=30\omega _{L}=1.5$,
respectively, and laser duration $\tau _{L}=5T_{L}$. 
 In addition, the XUV duration is an integer of the laser period, \textit{i.e.} $\tau_X=N T_L$.
In \Fref{figure3}(a) 
we have plotted the intracycle factor [equation \eref{F}] that is related to the angle-resolved PE spectrum considering 
%we consider
an XUV pulse of duration $\tau_{X}=T_{L}$, \textit{i.e.} $N=1$, with peak amplitude $F_{X0}=0.05$ and
%with no time delay, i.e., 
$\Delta _{X}=0$. 
In \Fref{figure4}, we show results for the ionization probability distribution for $%
F_{L0}=0.05$ in the left column (a, b, and c) and $F_{L0}=0.02$ in the right
column (d, e, and f). We see that the SCM electron yield [\Fref{figure4}(a)] is fully explained by the intracycle interference factor $F(\mathbf{k})$ [\Fref{figure3}(a)], the only difference between \Fref{figure3}(a) and
 \Fref{figure4}(a) is that in the latter the momentum distribution includes the factor  
$8\pi k\Gamma (k_{\perp})$ [see  equations \eref{dP} and (\ref{Tif4})]. 
In the second row (b and e), we show results of
the SFA and in the third row (c and f), the corresponding numerical solution
of the TDSE. Due to the close agreement between the SCM and the SFA with the
TDSE angle-resolved energy distributions, one may conclude that the effect
of the Coulomb potential on the energy spectrum is very small if not
negligible. However, the analysis of the effect of the Coulomb potential of
the remaining core on the electron yield deserves a thorough study, which is
beyond the scope of this paper.

%-------FIGURE 4----------------------------------------------------------
%-------FIGURE 4----------------------------------------------------------

For the laser and XUV parameters used in the left column of \Fref{figure4}, the
lower and upper classical boundaries of the energy distributions in the
direction along the polarization axis ($\theta=0^{o}$ and $\theta=180^{o})$ are $\left( v_{0}-F_{L0}/\omega
_{L}\right) ^{2}/2\simeq 0.086$ and $\left( v_{0}+F_{L0}/\omega _{L}\right)
^{2}/2\simeq 2.91,$ respectively \cite{Gramajo16}. The enhancement of the
probability distribution near threshold in the TDSE calculation in \Fref{figure4}(c)
is due to ATI ionization by the laser field (with no XUV pulse). This
contribution is highly suppressed in the SFA calculations \cite{Arbo08a} in
 \Fref{figure4}(b) and completely neglected in our SCM in \Fref{figure4}(a). 
For emission perpendicular to the polarization direction ($\theta=90^{o}$), the lower and upper classical
boundaries  (first column of \Fref{figure4}) are $\left[ v_{0}^{2}-(F_{L0}/%
\omega _{L})^{2}\right] /2=0.5$ and $v_{0}^{2}/2=1,$ respectively, for the case that $F_{L0}=0.05$ \cite{Gramajo17}. 
We see that the quantum SFA and TDSE results circumscribe to
the classical boundaries, except for a thin (in energy domain) decaying
probability beyond the classical boundaries. As shown in Sec. \ref{theory},
the SCM predicts %an optical-phase-dependent 
a discontinuity of the intracycle
stripes which, in the case of \Fref{figure4} %($\phi =0$),
is set at $E_{\mathrm{dis}}=v_{0}^{2}/2=1$, as is clearly observed in \Fref{figure4}(a) and \ref{figure4}(d)$.$ The
intracycle stripes for forward emission ($\theta <90^{\circ }$) have
positive slope at the left of the discontinuity ($E<E_{\mathrm{dis}}=1$), whereas they have negative slope at the right of it ($E>E_{%
\mathrm{dis}}=1$) in
\Fref{figure4}(a) and \ref{figure4}(d); and the opposite behavior for backward emission ($%
\theta >90^{\circ }$). 
We observe that such discontinuity is blurred in the quantum SFA and TDSE
calculations, where the two kind of intracycle stripes (with positive and
negative slope) coexist in an energy region close to $E_\mathrm{dis}$.

In order to study the dependence of the angle-resolved photoelectron
spectrum with the laser intensity, we show in the right column of \Fref{figure4} the
results using a laser peak field of $F_{L0}=0.02$. As the laser intensity is
lower than the used in the first column, the classically allowed region
shrinks. In particular, the energy distribution along the polarization axis is bounded by the lower $\left( v_{0}-F_{L0}/\omega _{L}\right)
^{2}/2\simeq 0.51$ and upper $\left( v_{0}+F_{L0}/\omega _{L}\right)
^{2}/2\simeq 1.64$ classical limits \cite{Gramajo16}. In turn, the classical
boundaries for emission perpendicular to the polarization axis are $\left[
v_{0}^{2}-(F_{L0}/\omega _{L})^{2}\right] /2\simeq 0.92$ and $v_{0}^{2}/2=1,$
being the last one insensitive to the laser intensity \cite{Gramajo17}. 
From Figures \ref{figure4}(a) and \ref{figure4}(b) 
we observe that the number of intracycle stripes diminishes
as the laser intensity decreases. For the TDSE calculations in \Fref{figure4}(f), we
observe a much lower contribution from near-threshold ATI by the laser
compared to \Fref{figure4}(c) since the intensity of the laser in the latter is
only the $16\%$ of the corresponding to the former. From a
direct comparison between the angle-resolved photoelectron spectra for
different laser intensities, we can conclude that they can be very useful at
the time of experimentally determining the elusive magnitude of the laser
intensity.

%-------FIGURE 5----------------------------------------------------------
\begin{figure}[h]
  \centering
    \includegraphics[width=0.7\textwidth]{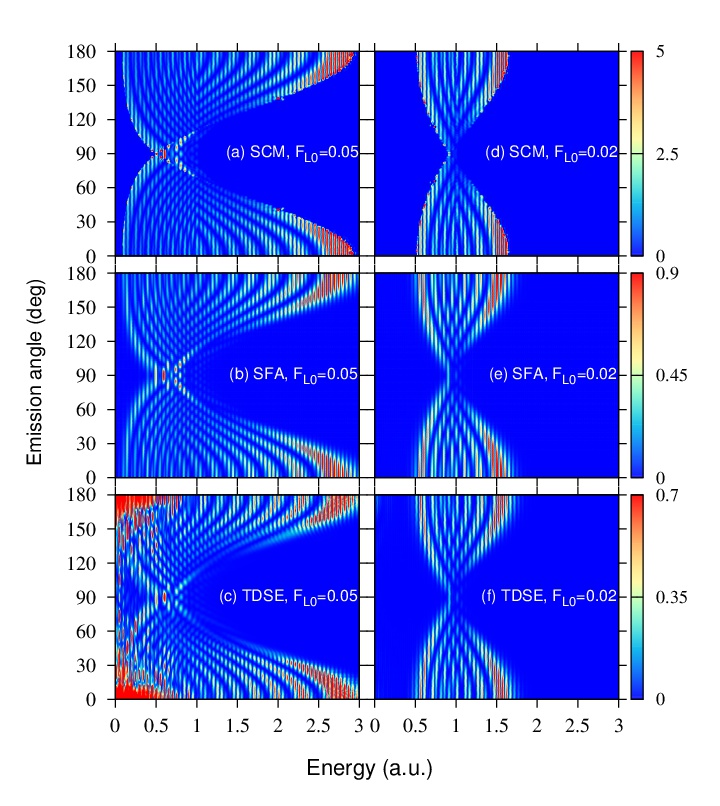}
  \caption{Angle-resolved photoelectron spectra in arbitrary units for an XUV pulse
duration of $\tau_{X}=2T_{L}$ and time delay $\Delta_{X}=0$, calculated at different laser peak fields ($F_{L0}=0.05$ a.u. in a, b, and c, and $F_{L0}=0.02$ a.u. in d, e, and f)
within the SCM (a and d), the SFA (b and e) and the TDSE (c and f). The IR
laser frequency is $\omega_{L}=0.05$ a.u. and the XUV pulse parameters are $%
\omega_{X}=1.5$ a.u. and $F_{X0}=0.05$ a.u..}
  \label{figure5}
\end{figure}
%-------FIGURE 5----------------------------------------------------------

As explained in Sec. \ref{theory}, for an XUV pulse of duration longer than
the laser period, intercycle interferences give rise to the formation of
sidebands. We clearly see the sideband formation in \Fref{figure5}, where the duration of the XUV
pulse involves two optical cycles, \textit{i.e.}, $\tau _{X}=2T_{L}.$ The rest of the
parameters are the same as in \Fref{figure4}.
 In general, we observe that the domain
of the angle-resolved energy distribution is independent of the XUV pulse duration ($\tau_{X} \geq T_{L}$) (the same as in \Fref{figure4}). The
only difference between the angle-resolved photoelectron spectra of \Fref{figure4}
and \Fref{figure5} is the formation of the sidebands, here depicted as vertical
isoenergetic lines at energy values $E_n$ according to equation (\ref{energy}) and separated by the photon energy $\omega _{L}=0.05$. We see that
the sidebands, stemming from the coherent superposition of the contributing
trajectories at the two different optical cycles, are modulated by the
intracycle pattern of \Fref{figure4}, due to the contributing trajectories within
the same optical pulse. So far, we have seen that the SCM and SFA predicted
backward-forward symmetrical emission, i.e., $\theta \leftrightarrow \pi
-\theta $ [see Figures \ref{figure4}(a), \ref{figure4}(b), \ref{figure4}(d), \ref{figure4}(e) and Figures \ref{figure5}(a), \ref{figure5}(b), \ref{figure5}(d), and \ref{figure5}(e)]. This
symmetry approximately holds but it is not exact in the TDSE calculations.
There are two reasons for the backward-forward symmetry breaking: The effect
of the Coulomb potential of the remaining ion and the depletion of the
ground state \cite{Arbo06a,Arbo08a,Nagele11,Arbo14}. These two effects are
completely neglected within the SFA and, therefore, also within the SCM.

%-------FIGURE 6----------------------------------------------------------
\begin{figure}[h]
  \centering
    \includegraphics[width=0.8\textwidth]{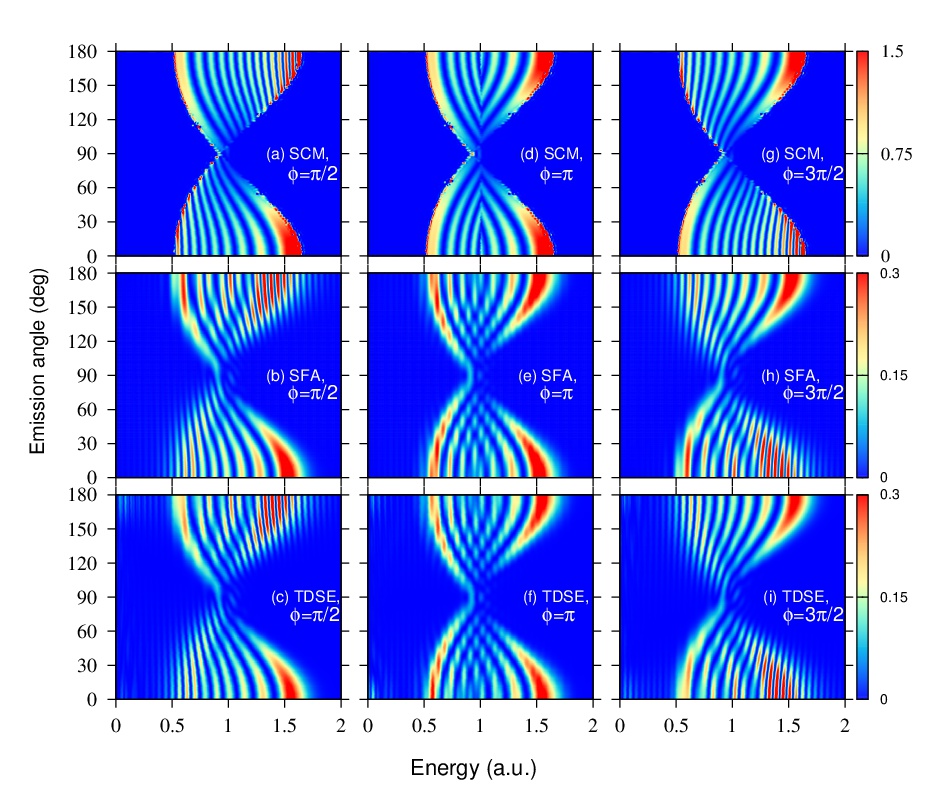}
  \caption{Angle-resolved photoelectron spectra in arbitrary units for an XUV pulse
duration of $\tau_{X}=T_{L}$ calculated at different optical phases ($%
\phi=\pi/2$ in a, b, and c, $\pi$ in d, e, and f, and $3\pi/2$ in g, h, and
i) within the SCM (a, d and g), the SFA (b, e and h) and the TDSE (c, f and
i). $F_{X0}=\omega_{L}=0.05$ a.u. $\omega_{X}=1.5$ a.u. and $F_{L0}=0.02$ a.u..}
  \label{figure6}
\end{figure}
%-------FIGURE 6----------------------------------------------------------

Another way of breaking the forward-backward symmetry is by including a time
delay $\Delta_X$ with respect to the hitherto XUV beginning time $\tau _{L}/2-\tau _{X}/2$ for the case of the two co-centered pulses.
For the sake of comparison, let us define the module $2\pi $
optical phase as the phase of the starting time of the XUV pulse with
respect to the vector potential $\mathbf{A}(t)$ as 
\footnote{Here the $2\pi $-equivalence $a\equiv b$ means that $(a-b)/2\pi $ is integer.} 
\begin{equation}
\phi \equiv \omega _{L}t_{Xb}=\omega _{L}\Delta _{X}+(N_{L}-N)\pi ,
\label{phase}
\end{equation}%
where $\phi$ is restricted to $0\leq \phi <2\pi $.
By varying $\Delta _{X}$ in equation (\ref%
{i-field}), the optical phase $\phi $ defined in equation (\ref{phase}) changes
accordingly. Whereas the active window for XUV ionization shifts in the time
domain, the vector potential of the laser pulse changes its shape relative
to the active window, with an ensuing change of shape of the intracycle
interference pattern. 
In \Fref{figure6}, we observe how the intracycle interference
pattern (angle-resolved photoelectron spectrum for $\tau _{X}=T_{L}$) changes
when the optical phase varies, i.e, $\phi =\pi /2,$ $\pi ,$ and $3\pi /2$,
for the left (a, b, and c), middle (d, e, and f), and right (g, h, i)
columns, respectively. In Figures \ref{figure6}(a), \ref{figure6}(d), and \ref{figure6}(g) the SCM exhibits the change of the intracycle interference pattern with $\phi $. 
For the optical phase $\phi =\pi $ in \Fref{figure6}(d) the active window is shifted by half laser period with respect of the optical phase $\phi =0$ in \Fref{figure4}(d)
and, thus, the vector potential relative to the active window inverts (it
changes sign). 
Therefore, we should expect a forward-backward inversion of
the angle-resolved spectrum, however, due to its forward-backward symmetry
the electron emission stays unaltered. The forward-backward inversion can be
observed by comparing Figures \ref{figure6}(g) and \ref{figure6}(a) since the change of the optical phase is $\pi$ [equation \eref{phase}].
We note that, similarly to the $\phi=0$ case, the aforementioned discontinuity occurs at $E_\mathrm{dis} = 1$ in \Fref{figure6}(d). 
Neverthless, for the general case, the line of discontinuity depends on $\phi$ (through $t_{Xb}$) and the emission angle; in fact, it is possible to deduce that the energy values where the discontinuity take place follow 
\begin{equation}
E_{\mathrm{dis}}(t_{Xb},\theta)=\frac{1}{2}\left[\sqrt{v^{2}_{0}-A^{2}_{L}(t_{Xb})\sin^{2}\theta}-A_{L}(t_{Xb})\cos{\theta}\right]^{2}.
\label{Edisc}
\end{equation}
This equation generalizes the discontinuity  previously deduced for forward and perpendicular cases \cite{Gramajo16, Gramajo17}.
For $\phi=0$ and $\phi=\pi$, $E_\mathrm{dis}=v^{2}_{0}/2=1$a.u., which is independent of the emission angle $\theta$. For $\phi =\pi /2$, \Fref{figure6}(a), the
discontinuity has displaced to one classical boundary in equation
(\ref{Edisc}), whereas for $\phi =3\pi /2$, \Fref{figure6}(g), the discontinuity coincides with the
other classical boundary in equation (\ref{Edisc}), loosing its entity in both cases.
In the supplemental material, we show a movie of how the SCM
angle-resolved photoelectron spectrum changes with the optical phase for a larger
number of optical phases than the depicted in \Fref{figure6}. There it is easy to observe the angle dependence of $E_\mathrm{dis}$.
The SFA and TDSE angle-energy distributions in the respective
Figures \ref{figure6}(e) and \ref{figure6}(f) blur the mentioned discontinuity as previously
discussed for $\phi =0$. It is worth to mention that the SCM resembles SFA and TDSE
angle-resolved photoelectron distributions quite accurately for all optical
phases.

%-------FIGURE 7----------------------------------------------------------
\begin{figure}[h]
  \centering
    \includegraphics[width=0.8\textwidth]{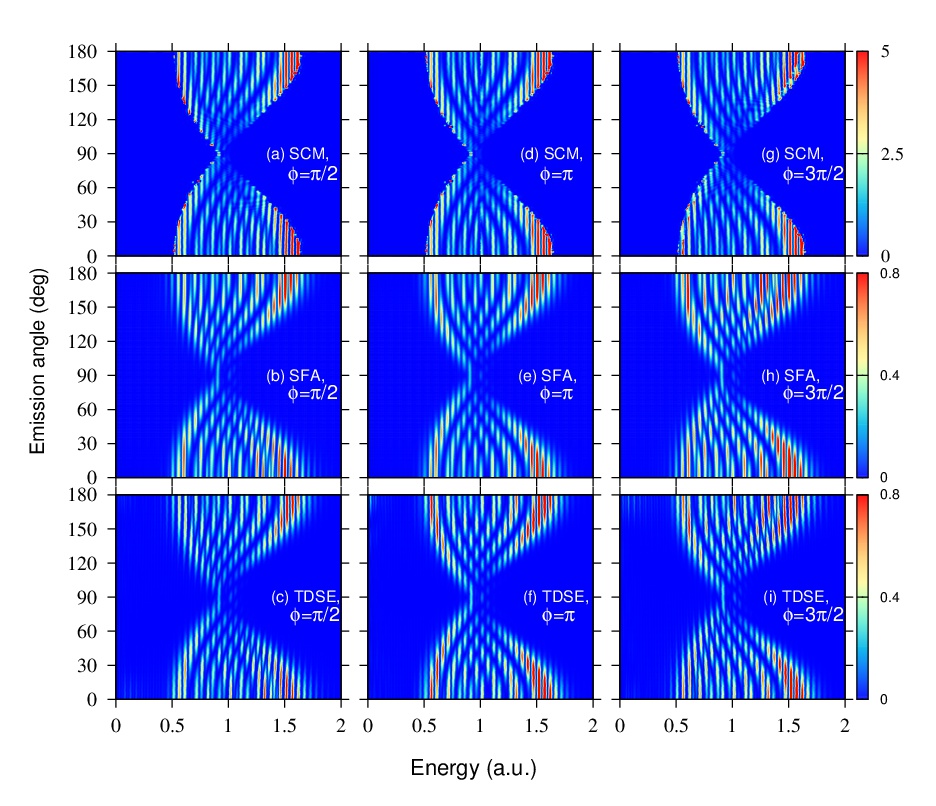}
  \caption{Angle-resolved photoelectron spectra in arbitrary units for an XUV pulse
duration of $\tau_{X}=2 T_{L}$ calculated at different optical phases ($%
\phi=\pi/2$ in a, b, and c, $\pi$ in d, e, and f, and $3\pi/2$ in g, h, and
i) within the SCM (a, d and g), the SFA (b, e and h) and the TDSE (c, f and
i). The IR laser parameters are the same as in the \Fref{figure6}.}
  \label{figure7}
\end{figure}
%-------FIGURE 7----------------------------------------------------------

In order to study the formations of sidebands for different optical phases $\phi$, in  \Fref{figure7} we plot the angle-resolved
photoelectron spectrum for the same XUV and laser parameters as in \Fref{figure6} except that the XUV pulse duration is $\tau _{X}=2T_{L}.$ We
see the presence of the sidebands produced by the intercycle interference
between the contributions of the photoemission within the first and second optical cycles. As shown before, sidebands (the intercycle pattern) are modulated by the intracycle pattern of  \Fref{figure6}. As explained in the last paragraph, we see that the
energy-resolved photoelectron spectrum is exactly symmetric when calculated
for $\phi =\pi $ within the SCM [\Fref{figure7}(d)] and the SFA [\Fref{figure7}(e)], and
approximately symmetric when calculated within the TDSE. Besides, we see
that the asymmetry observed in the intracycle interference for
optical phase $\phi =\pi /2$ (a, b, and c) and $3\pi /2$ (g, h, and i) is
strongly suppressed in the respective figure \ref{figure7}(a, b, and c) and figure \ref{figure7} (g, h, and i) compared to  \Fref{figure6}, due to
the presence of the intercycle interference in the former.
We see
that the dependence of the PE spectra on the delay diminishes as the XUV
duration increases.
In fact,  in the limit of infinite durations,  the sidebands are represented as delta functions in the energy domain,
i.e., $\delta (E- E_{n})$ where $E_{n}$ is given by equation (\ref{energy}),
independently of the XUV delay in agreement with the SPA.

%-------FIGURE 8----------------------------------------------------------
\begin{figure}[h]
  \centering
    \includegraphics[width=0.7\textwidth]{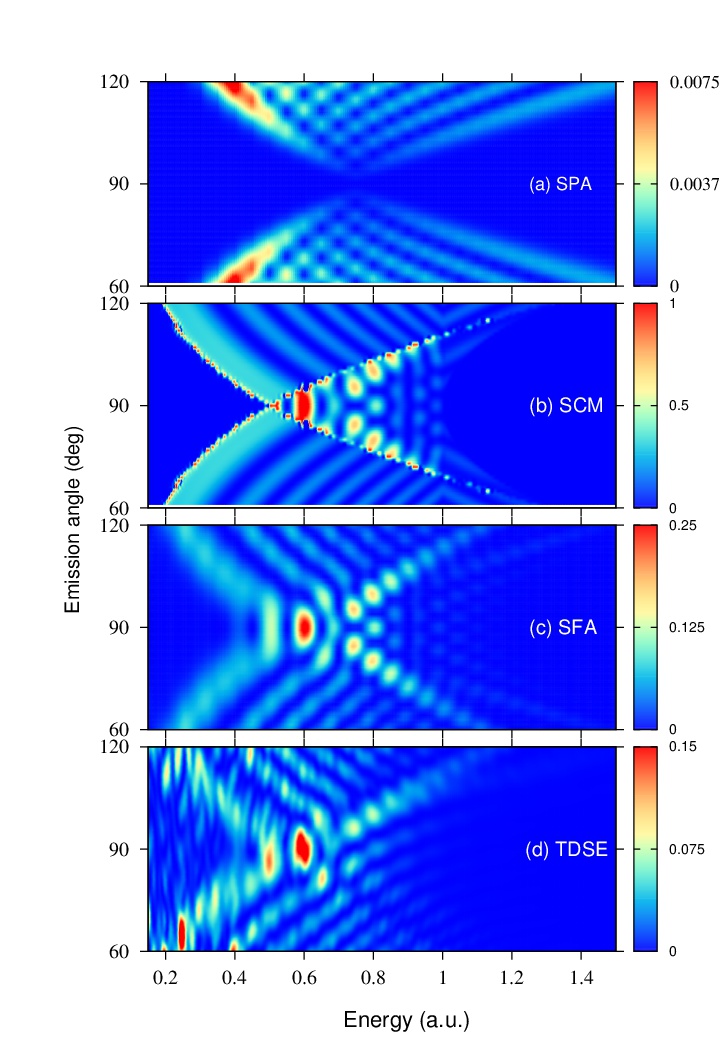}
  \caption{Angle-resolved
PE spectra in atomic units in the perpendicular-like region
% Perpendicular-like region in the angle-resolved photoelectron spectra 
for an XUV pulse duration of $\tau_{X}=T_{L}$
calculated within the SPA (a), the SCM (b), the SFA (c), and the TDSE (d). The IR laser
parameters are the same as in the figure \ref{figure4} and $F_{L0}=0.05$a.u..}
  \label{figure8}
\end{figure} 
%-------FIGURE 8----------------------------------------------------------

The SPA has been widely employed to depict satisfactorily experimental results %as in 
\cite{Meyer2006, Meyer2008, Meyer2010JPB, OKeeffe04, Dusterer2016,Hayden16}.
However, since its dipole element involved is proportional to $\mathbf{e}_{z}\cdot \mathbf{k}=k_{z}$, the
SPA predicts no emission in the direction perpendicular to the polarization axis of the laser field  \cite{Maquet07,Jimenez13}. 
For that, in order to compare  the emission yield in the perpendicular direction for different theories, we focus
%
%Now we want to focus 
on the intracycle interference pattern in the
perpendicular-like region. 
The Figures \ref{figure8}(a, b, c and d) are an augmentation of
the angle-resolved photoelectron spectrum near the perpendicular-like region
for the same XUV and laser parameters used hitherto for the high laser
intensity, i.e., $F_{L0}=0.05$ and $\tau _{X}=T_{L}$ in \Fref{figure4}.
Here we note that the SPA predicts discrete final energy values according to the sideband values, since it is derived for infinitely long pulses [see for example Equation (2.10) of  \cite{Jimenez13}]. 
Not only have we included the ponderomotive shift, so that the  sideband positions are in agreement to the energy conservation equation \eref{energy}, but  we have also extended it linearly for continuous energy values 
for better comparison with other theories. In this way, SPA can be  interpreted as the modulator of the sidebands, i.e. the intracycle pattern.
We observe a qualitative agreement among the SCM (b), SFA
(c), and TDSE (d) distributions, as discussed previously. However, the SPA model [\Fref{figure8}(a)] is shifted towards higher energy values. More importantly, the SPA exhibits null electron perpendicular emission, according to Ref. \cite{Jimenez13}. The TDSE angle-resolved photoelectron spectrum shows some degree of forward-backward asymmetry since the Coulomb force of the remaining core cannot be neglected. This fact can be easily understood since the force of the electric field is weak (and vanishes in the perpendicular direction) and, therefore, the main hypothesis of the SFA fails close to perpendicular emission ($\theta \sim 90^{\circ }$). 
%
%We observe that the SPA angle-resolved energy distribution shows a displacement to high energy values respect to the SCM and the quantum results (SFA- TDSE). 

%-------FIGURE 9----------------------------------------------------------
\begin{figure}[h]
  \centering
    \includegraphics[width=0.7\textwidth]{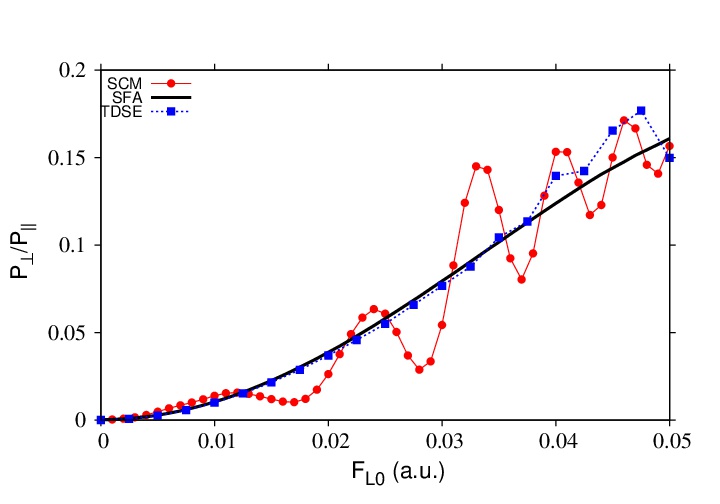}
  \caption{Ratio of transversal and forward total emission probabilities as function of laser amplitude within the SCM (red solid line with circles), SFA (black solid line) and the TDSE (blue dashed line with squares).}
  \label{figure9}
\end{figure} 
%-------FIGURE 9----------------------------------------------------------

As observed in \Fref{figure4}, the emission is highly dependent on the laser intensity. 
Because of this, it is worth to analyze the contribution of the ionization probability in the perpendicular direction for different intensities. 
In the following, we compute the total ionization probability at fixed emission angle, integrating in energy the PE spectrum in energy  equation (\ref{dP}).
In \Fref{figure9} we present the ratio of perpendicular and forward emission, i.e., $P_\perp/P_\parallel$, as a function of the laser electric field amplitude $F_{L0}$ 
within the SCM, the SFA and the TDSE approaches.  We have considered fixed $F_{X0}=0.05$, since the SFA and SCM probabilities are proportional to the XUV intensity we expect the result does not change for different values of $F_{X0}$. As we have observed before in \Fref{figure4}, the TDSE spectra show a high emission probability at low energies due to the ATI ionization by the IR pulse alone. Therefore, in order to compute the TDSE total ionization probability at $\theta=0^{o}$ we have omitted the contribution of direct ionization probability, i.e. $P_\parallel \simeq P_\parallel^{ \scriptstyle XUV+IR}-P_\parallel^{\scriptstyle IR}$. 
Then, the TDSE ratio be comes very sensitive to this straightforward estimation of $P_\parallel$, especially for $F_{L0} \ge 0.04 $, as we can observe in \Fref{figure9}.
%Then, the TDSE ratio shows structures in the region $0.04 \le F_{L0}\le0.05$ due to this straightforward $P_\parallel$ estimation.
%
The three theories predict  that $P_\perp/P_\parallel$ increase with the laser amplitude, showing that for higher laser intensities the perpendicular emission be comes significant and cannot be neglected. The SCM ratio shows oscillations around the quantum calculations (TDSE and SFA), which are related to the abrupt cut in the energy domain due to the classical boundaries.
%

%---------------------------------------------------

\section{\label{conc}Conclusions}

%---------------------------------------------------

We have studied the electron spectrum at all emission angles produced by atomic hydrogen initially in the ground state subject to an XUV pulse in the presence of an infrared laser pulse. We have generalized the SCM previously posed to study LAPE in the direction along the polarization axis \cite{Gramajo16} and
perpendicularly to it \cite{Gramajo17}. The classically allowed angle-energy
domain can be divided in two different regions: The parallel-like and the
perpendicular-like regions. In the former, two classical electron trajectories per optical cycle contribute to the (intracycle) interference pattern which modulates the sidebands stemming from the (intercycle) interference of the electron trajectories at different optical cycles. In the latter, the four contributing classical electron trajectories can be grouped in two pairs in one optical cycle, giving rise to  grosser (intrahalfcycle) structure which modulates the intracycle pattern. We have shown that, as the laser intensity increases, the angle-resolved photoelectron spectra become wider in the energy domain showing a considerable extended perpendicular-like region bounded within the classical domain. We have observed a very good agreement between the SCM angle-resolved energy spectrum with the corresponding SFA and the \textit{ab
initio} calculations of the TDSE. The relevance of the SCM is
evident for emission in the perpendicular-like region. Whereas the
SPA \cite{Maquet07,Jimenez13} and Kazansky's
first-order time-dependent perturbation theory \cite%
{Kazansky10a,Kazansky10b,Kazansky12,Kazansky14} predict null perpendicular
emission for ionization from an $s$ state, our SCM foresees apreciable non-zero
probability in the perpendicular-like region in the line of Bivona's
theory \cite{Bivona10} and in agreement with SFA and TDSE calculations. The TDSE emission yield
experiences a breaking of the forward-backward symmetry for short XUV pulses, which is mostly
recovered as the XUV pulse duration comprises a few laser optical cycles.
Finally, we have analyzed the angle-resolved electron spectrum for different
time delays $\Delta _{X}$ between the two pulses. We have also shown that
when the XUV pulse duration is a multiple of the laser period and for
optical phases $\phi =0$, and $\pi$, the emission within the TDSE is highly
symmetrical in the forward and backward direction, in agreement with the SFA
and SCM forward-backward. Forward-backward asymmetries come up for different, from $\phi=0$ and $\pi$, optical phases.

To conclude, we point out that the observation of the aforementioned results should be attainable with the current experimental performance. 
In particular, the angle-resolved photoelectron spectra come accessible  using, for example, velocity map imaging spectometer or COLTRIMS technique with long term stability of the synchroization between the XUV and IR fields \cite{Duesterer13,Haber10,Picard14}.
 We think that experimental measurements with strong lasers would be highly desirable in order to corroborate the rich structure of the PE angle-resolved spectra specially in the perpendicular region.

%\begin{acknowledgments}

%---------------------------------------------------

\section*{Appendix. Ionization times and transition matrix calculations}

%---------------------------------------------------

\begin{figure*}[h!]
%\centering
\subfigure[ Parallel-like case]{\includegraphics[width=0.5\textwidth]{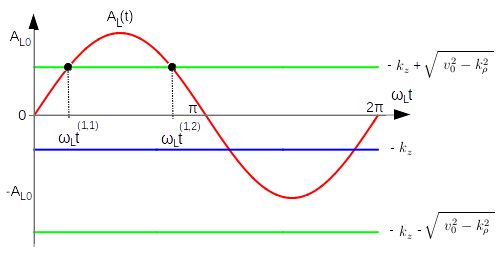}}%\\
\subfigure[ Perpendicular-like case]{\includegraphics[width=0.5\textwidth]{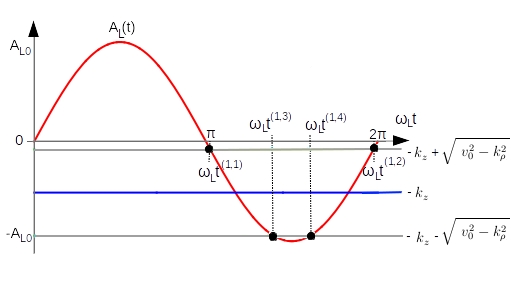}}
\caption{Emission times solutions of equation (\ref{circle}) as intersection of the three curves, $A_{L}(t)=A_{L0}\sin{(\omega_{L}t)}$ in red solid line and the constants $\beta_{\pm}=-k_{z}\pm \sqrt{v_{o}^2-k_{\perp}^2}$ %in green and light gray solid line, 
for a fixed electron momentum with $k_{z}>0$ and in the first IR oscillation cycle. 
In this scheme, the beginning time of the XUV pulse is zero.
(a) Parallel-like case characterized by two-ionization times per optical cycle.
(b) Perpendicular-like case characterized by four-ionization times per optical cycle. 
}
\label{figure2}
\end{figure*}

The ionization times $t_{s}$ that fulfill equation \eref{circle}  are calculated as the intersection of the horizontal lines 
\begin{equation}
\beta_\pm(\mathbf{k}) \equiv -k_z \pm \sqrt{v_0^2 -k_\perp^2} \label{betamm}
\end{equation} 
and the vector potential $A_{L}(t)$.
In figure \ref{figure2} we represent schematically how the ionization times $t_s$ are determined for a fixed momentum $\mathbf{k}$ with $k_z > 0$ (negative $k_z$ can be straightforward deduced). 
We can distinguish two different situations depending if there are (or not) solutions with the negative branch of the square root.  When $-k_z - \sqrt{v_0^2 -k_\perp^2} = \beta_- (\mathbf{k})< - A_{L0}$ the negative branch never reaches any value of $A_L(t)$, and then there are only two times in the $j$th-optical cycle, i.e.,  $ t^{(j,1)}$ and $ t^{(j,2)}$.  This case, illustrated in \Fref{figure2}(a), corresponds to the called \textit{parallel-like} region. 
Under this condition, when $\beta_+(\mathbf{k}) > 0$ the emission times remain in the first half of the optical cycle.  As $\beta_+(\mathbf{k})$ decreases  to zero, the release time $t^{(j,1)}$ goes to the beginning of the laser cycle whereas the late release time $t^{(j,2)}$ goes to the middle of it [see \Fref{figure2}(a)]. Finally, when $\beta_+(\mathbf{k}) < 0$, the two ionization times move to the second half of the optical cycle.
This transition produces a discontinuity in the PE spectra as the one discussed in Ref. \cite{Gramajo16}.
On the other hand, 
when $\beta_- (\mathbf{k})> - A_{L0}$  
 the negative branch intersects $A_L(t)$ at  times $ t^{(j,3)}$ and $ t^{(j,4)}$ and, thus, ,  there are four ionization times per optical cycle [see \Fref{figure2}(b)]. 
This condition defines the \textit{perpendicular-like} region.
As before, 
the times $ t^{(j,1)}$ and $ t^{(j,2)}$ may be in the first or second half of the $j$th-cycle depending on the momentum value, whereas $ t^{(j,3)}$ and $ t^{(j,4)}$ are always in the second one.

%-------FIGURE 2----------------------------------------------------------
%-------FIGURE 2----------------------------------------------------------

The ionization times of different cycles are simply related to the first one through
\begin{equation}
t^{(j,\alpha)}=t^{(1,\alpha)}+(2\pi/\omega_{L})(j-1) \label{tj},
\end{equation}
 where $j =1,2$,...$N$ indicates the $j$th optical cycle, $N$ is the total number of laser cycles and $\alpha=1,2,3$ and $4$ correspond the four ionization times per cycle described before.
The solutions of equation \eref{circle} with $\beta_+(\mathbf{k})\geq 0$ lie in the first half cycle:
\begin{eqnarray}
t^{(1,1)}=\frac{1}{\omega_{L}} \sin^{-1}{\left|\frac{\beta_+(\mathbf{k})}{A_{L0} }\right|}
\qquad \textrm{and}\qquad
t^{(1,2)}=\frac{\pi}{\omega_{L}}- t^{(1,1)}.\label{para1}
\end{eqnarray}
Instead, if $\beta_+(\mathbf{k})<0$, they are in the second half cycle:
\begin{eqnarray}
t^{(1,1)} = \frac{\pi}{\omega_{L}}+\frac{1}{\omega_{L}}\sin^{-1}{ \left| \frac{\beta_+(\mathbf{k})}
{A_{L0}}\right| }
\qquad \textrm{and}\qquad
t^{(1,2)}=\frac{3\pi}{\omega_{L}}- t^{(1,1)}. \label{para2}
\end{eqnarray}
Furthermore, in the perpendicular-like region,  the  third and fourth ionization times are
\begin{eqnarray}
t^{(1,3)}=\frac{\pi}{\omega_{L}}+\frac{1}{\omega_{L}}
\sin^{-1}{  \left| \frac{\beta_-(\mathbf{k}) }{A_{L0}} \right|   } 
\qquad \textrm{and}\qquad
t^{(1,4)}&=&\frac{3\pi}{\omega_{L}}- t^{(1,3)}. \label{para3}
\end{eqnarray}

In general, 
the ionization times depend on the starting time $t_{Xb}$ of the active window
[see equation \eref{i-field}]. The previous analysis has been done for an IR laser whose vector potential vanishes at $t_{Xb}$. When this is not the case, we have to considerer a shift in the ionization times equations (\ref{para1}) to (\ref{para3}).

Finally, the transition matrix of equation (\ref{Tifsaddle}) considering two or four ionization times per IR cycle in the parallel- or perpendicular-like situations  results in 
\begin{eqnarray}
T_{if}&=& \sum_{j=1}^{N}\left[ 
\sum_{\alpha=1}^{2}\, g(\mathbf{k},t^{(j,\alpha)})   \quad
+  \quad \Theta{ \left(1+\frac{\beta_-(\mathbf{k})}{A_{L0}}\right)}
\sum_{\alpha=3}^{4}\, g(\mathbf{k},t^{(j,\alpha)})
\right]
\label{Tif2} ,
\end{eqnarray}%
where we have introduced  the Heaviside function $\Theta$ so that
 the second term contributes only in the perpendicular-like case $\beta_{-}(\mathbf{k})\geq-A_{L0}$. 
According to equation \eref{Tifsaddle} the terms to add are
\[
g(\mathbf{k},t^{(j,\alpha)})=\frac{\sqrt{2\pi }F_{X0}(t^{(j,\alpha)})d_{z}(\mathbf{k}+\mathbf{A}_{L}(t^{(j,\alpha)}))}{|\left[ \mathbf{k}+\mathbf{A}_{L}(t^{(j,\alpha)}) \right]\cdot \mathbf{F}_{L}(t^{(j,\alpha)})|^{1/2}}\, \exp \left[\rmi S(t^{(j,\alpha)})%
+ \rmi \frac{\pi}{4} \mathrm{sgn} [ \ddot{S}(t^{(j,\alpha)}) ] \right].
\]
We note that the $z$-component of the dipole element [equation \eref{dip}] is proportional to $k_z+ A_L(t^{(j,\alpha)})$, thus $d_z$ evaluated in the first or second IR halfcycle has equal magnitude and opposite sign. 
Therefore,  we  factorize $|d_z|$ and introduce a minus sign ahead the Heaviside function in the second term of equation \eref{Tif2}. 

To evaluate the action $S$ [equation \eref{action}] at the ionization times, let us considerer the accumulated action 
$\Delta S_{j\pm}=S(t^{(j,\alpha)})-S(t^{(j,\alpha+1)})$
and the action average
$
\bar{S}_{j\pm}=[S(t^{(j,\alpha)})+S(t^{(j,\alpha+1)})]/2
$
of two trajectories released in the same $j$th cycle, 
where the sign $+$($-$) corresponds to $\alpha =1$(3) respectively.
This results in 
$\exp \left[\rmi S(t^{(j,\alpha  )})\right]=\exp ( \rmi \Delta S_{j\pm}/2 + \rmi \bar{S}_{j\pm})$ and
$\exp \left[\rmi S(t^{(j,\alpha+1)})\right]=\exp (-\rmi \Delta S_{j\pm}/2 + \rmi \bar{S}_{j\pm})$.
Replacing Equations (\ref{tj}) to (\ref{para3}) into (\ref{action}) we found that both accumulated actions are independent of the cycle $j$,
\begin{eqnarray}
\Delta{S}_{\pm}&=\frac{\tilde{S}}{2}
\left[ \frac{2}{\pi}\sin^{-1}\left|\beta_{\pm}(\mathbf{k})/A_{L0} \right|-1\right]-\frac{\textrm{sgn} [ \beta_{\pm}(\mathbf{k})  ]}{2\omega_{L}}\left(4k_{z}+\beta_{\pm}(\mathbf{k})   \right)\sqrt{A_{L0}^2-\beta_{\pm}^2(\mathbf{k})} \nonumber%
\\  \label{DeltaSpara}
\end{eqnarray}
and $ \bar{S}_{j\pm}= S_{0\pm} + j \tilde{S}$  depends linearly on the cycle index $j$, where 
$\tilde{S} = (2\pi/\omega_{L} )(E + I_{p} + U_{p}-\omega_{X})$ and $S_{0-}=3 S_{0+}=-3\tilde{S}/4$.

After a bit of algebra it can be shown that each of the $N$ terms in equation (\ref{Tif2}) is proportional to
\[ 
 e^{\rmi j \tilde{S}} 
\left[
f_+(\mathbf{\scriptstyle k})
\cos{  \left(\frac{\Delta S_+ }{2} +  \frac{\pi}{4} \mathrm{sgn}[\beta_+(\mathbf{k})]  \right) } 
 -
 \Theta{  \scriptstyle \left(1+\frac{\beta_-}{A_{L0}}\right)}
f_-(\mathbf{\scriptstyle k})
\cos{\left(\frac{\Delta{S}_{-}}{2}
-  \frac{\pi}{4}\mathrm{sgn}[\beta_{-}(\mathbf{k})] \right) }
\right],
\]
where 
\begin{equation}
f_\pm(\mathbf{k}) = e^{\pm\rmi\tilde{S}/4} \, | 1 -(\beta_{\pm}( \mathbf{k})/A_{L0})^{2}|^{-1/4}.  
\label{ff}
\end{equation}
Finally, when the sum over the $N$ optical cycles is achieved,  the emission probability results in equation (\ref{Tif4}).
The precedent results have been deduced for the $k_z\ge 0$, however  the negative cases can be straightforwardly deduced replacing $k_z$ by $|k_z|$ in previous equations.

\ack
Work supported by CONICET PIP0386, PICT-2012-3004 and PICT-2014-2363 of
ANPCyT (Argentina), and the University of Buenos Aires (UBACyT
20020130100617BA).
%\end{acknowledgments}

%---------------------------------------------------
%\bibliographystyle{plain}
\bibliographystyle{iopart-num}
\bibliography{biblioA}

% Produces the bibliography via BibTeX.
%---------------------------------------------------

\end{document}